\documentclass[11pt,a4paper]{article}
\usepackage{jheppub,amsmath,amssymb,epsfig,graphics,graphicx,dsfont}

\usepackage{hyperref}

\newcommand{\tsty}{\textstyle}

\def\Sl#1{\rlap{\hbox{$\mskip 1 mu /$}}#1}
\def\I{{\mathds 1}}

\title{Mass Generation and Related Issues from Exotic Higher Dimensions}

\author[a]{M. Rojas}
\author[b,c]{M.A. De Andrade}
\author[d,c]{L.P. Colatto}
\author[e]{J.L. Matheus-Valle}
\author[f]{L.P.G. De Assis}
\author[f,c]{J.A. Helay\"{e}l-Neto}
\affiliation[a]{ DEX, Universidade Federal de Lavras, CEP 37200-000, MG, Brasil}
\affiliation[b]{Universidade do Estado do Rio de Janeiro, Resende, CEP 27537-000, RJ, Brasil}
\affiliation[c]{GFT-JLL Petr\'opolis, CEP 25651-000, Brasil}
\affiliation[d]{CEFET-RJ UnED-Petr\'opolis, CEP 25620-003, RJ, Brasil}
\affiliation[e]{DF-ICE, Universidade Federal de Juiz de Fora, CEP 36036-330, MG, Brasil}
\affiliation[f]{CBPF-LAFEX, Urca, Rio de Janeiro, CEP 22290-180, RJ, Brasil}

\emailAdd{mrojas@cbpf.br} 
\emailAdd{deandrade.marco@gmail.com}
\emailAdd{colatto@cbpf.br} 
\emailAdd{zeluiz@fisica.ufjf.br}
\emailAdd{lpgassis@cbpf.br}
\emailAdd{helayel@cbpf.br}

\abstract{ 
\noindent The main purpose of this work is to show that massless Dirac equation formulated for non-interacting Majorana-Weyl spinors in higher dimensions, particularly  in $D = 1 + 9$ and $D = 5 + 5$, can lead to an interpretation of massive Majorana and Dirac spinors in $D=1+3$.
By adopting suitable representations of the Dirac matrices in higher dimensions, we pursue the investigation of which higher dimensional space-times and which mass-shell relation concerning massless Dirac equations in higher dimensions may induce massive spinors in $D=1+3$.  The mixing of the chiral fermions in higher dimensions may induce a mechanism such that four massive Majorana fermions may show up and, at an
appropriate limit an almost zero and a huge mass show up with corresponding left-handed and right-handed eigenstates. This mechanism, in a peculiar way, could reassess the See-Saw scheme associated to neutrino with Majorana-type masses. Remarkably the masses of the particles are fixed by the dimension decoupling/reduction scheme based on the mass Lorentz invariant term, where one set of the decoupled dimensions are the ``target'' coordinates frame and the other set of coordinates is the composing block of the mass term in lower dimensions. 
This proposal should allow us to understand the generation of hierarchies, such as the fourth generation, for the fermionic masses in $D=1+3$, or in lower dimensions in general, starting from the constraints between the energy and the momentum in $D=n+ n$. For the initial $D=5+5$ Majorana-Weyl spinors framework using the Weyl representation to the Dirac matrices we observe an intriguing decomposition of space-time that result in two very equivalent $D=1+4$ massive spinors which mass term, in $D=1+3$ included, is originated from the remained/decoupled component and that could induce a Brane-World mechanism.  
}

\begin{document}

\maketitle

\section{Introduction}

Theories on space-times higher than the well-known $D=1+3$ have largely been studied in many aspects\cite{moretime}. Pure spatial additional dimensions could be interpreted as a geometrical extension of the $D=1+3$ and it has been extensively treated in the literature \cite{GSW,Polchinski,Zwiebach}.  Additional time directions pose a number of non-trivial questions, both at classical and quantum-mechanical levels. At the quantum level, these time directions mean that we are going to deal with negative norm square states on the Hilbert space, or shorter ghosts. At a geometric level, it implies we have to deal with non-compact little groups\cite{duff}. Nevertheless, developments have been done to attempt to understand the possible consequences and reflexes of additional time directions in the ordinary  $D=1+3$ theory\cite{moretime,duff}. It is remarkable Nambu's work\cite{nambu} where the canonical momentum of a particle evolving with two (time-like) parameters was reassess in an extended matrix version. It also includes potentials (as K\"ahler) in the classical mechanics Lagrangian and promotes a differential one to two-form in the Hamilton-Jacobi formalism. 

So the extensions of the standard model could shed light on some phenomena which involves mass term, such as the small mass of the neutrinos for instance. In this work, we apply a mass generation mechanism without the need of introducing a Higgs-type scalar field. This mechanism is based on a simple decomposition of the space-time in higher dimensions than four, using suitable Lorentz invariant energy-momentum relations. This approach was been proposed in a series of works \cite{gamma,RAHA}. We remark that Viollier \textit{et al}\cite{vio} some years ago have obtained similar results but to different aim. To our aim we start from massless fields in some space-time signatures to arrive to massive ones in four dimensions.  
Particularly, we are going to apply this mechanism to free Dirac Lagrangeans \cite{dirac} in ten-dimensi\-onal space-time which, based on the supersymmetry constraint, can have the signatures $D=1+9$, $D=9+1$ and $D=5+5$. The first two space-times are correlated by a change from space to time (and vice-versa) status. Because they do have essentially only one time-like direction they are the most exploited ones. This number prevents them to deal with negative norm states of the Hilbert space, or shortly ghosts.  Furthermore it implies that the compactify curled space is Euclidean, and so it is only geometrical (no dynamics). In fact, the compactification of time directions result in ghost degrees of freedom, which are complicated to be consistently treated. On the other hand, though the third signature $D=5+5$ present all the problems already pointed out, it can be connected to $D=2+2$ space-time in the string dynamics what has been discussed by Ketov, Nishino and Gates in the work of \cite{gates} for instance. And the $D=2+2$ case (that could have the origin on the $N=2$ strings) can be consistently treated, in a $N=1$ supersymmetric case, doing a truncations of the unphysical degrees of freedom maintaining the $N=1$ supersymmetry\cite{osw}. To this space-time ($D=5+5$), in another approach proposed by Hull\cite{Hull2}, it was claimed that a time-like compactification could not represent any inconsistency.  

It was verified that from hidden directions of a higher-dimensional background: When we fill this background with massless Majorana modes and break it into two parts \cite{gamma}, such that one of the parts is the $D=1+3$ space-time, we naturally get mass and we obtain results which resemble those appearing in the context of the See-Saw Mechanism \cite{RAHA}. In the present contribution, we extended this approach adopting model with $D=1+9$ (or $D=9+1$), and $D=5+5$ Minkowskian as our starting higher-dimensional space-times. We also consider the possibility to apply this scheme as a mechanism for neutrino mass generation (oscillation). In particular, our proposal fits into a series of works that review the potential use of models with extra time dimensions \cite{foster,Lanciani,Boyling,Patty}.

We also point out that our work sets out to motivate, on more geometrical and group-theoretic grounds, the revival of interest in a fourth generation of chiral matter, specially the fourth-generation neutrino. In the works of Ref. \cite{HHHMSU}, and a number of references quoted therein, it turns out the electroweak precision data do not exclude a fourth generation of quarks, charged leptons and their corresponding left and right neutrinos \cite{KPST}.
Indeed, from the higher-dimensional spaces we contemplate here, there naturally appears a fourth fermion generation. It is not our goal right now to present a phenomenological model with a strongly-coupled fourth generation; we intend to highlight that a dimensional compactification mechanism may yield the existence of a fourth-generation neutrino with a given mass hierarchy to be respected. Though we should say that there still lacks a fundamental interaction responsible for the condensation of a fourth-generation fermion that triggers the electroweak symmetry breaking, we believe that extensions of the Standard Model with a heavy fourth generation is a very appealing idea to dynamically justify the electroweak breaking \cite{BHL,BDM,BDEM}. This fact strongly motivates our work and the era opened up by the LHC is very appropriate to reassess the Standard Model beyond 3 generations of chiral matter. 

The outline of our work is as follows: in Section 2, we generally present the reduction scheme to go from $10$ to $4$ dimensions, without any commitment with the space-time signature.

Next, in Section 3, we focus on the $D=1+9$ case and contemplate its reduction to $D=1+3$. In Section 4, we consider the $D=5+5$ space-time and we exploit its features in contrast with the $D=1+9$ case. Specially, we present what we refer to as the intermediate cases: $D=1+4$ and $D=2+3$. Finally, in Section 5, we cast our Concluding comments and we highlight some potential applications of our procedure to produce different generations out of dimensional reduction.

Helpful details are all cast in the Appendix and its sub-sections.

\section{From \texorpdfstring{$10D$}~ to \texorpdfstring{$4D$}~ Spinor Fields}

First of all, we are going to settle down the notation to the construction of the $D=1+9$ and $D=5+5$ spinor models \cite{gamma,RAHA}.
In order to define the ``target" standard model that can be represented by ordinary four-dimensional massive spinor fields, we recall the algebra of the proper orthochronous Lorentz group SO$^+$(1,3) which has the metric $\eta^{\mu\nu} = {\rm diag}(+---)$ is 
\begin{equation}
\{\gamma^\mu, \gamma^\nu \} = 2\eta^{\mu\nu}\I
\label{diracalg}
\end{equation}
where the index $\mu=(0, ..., 3)$ labeled the tensor representation SO$^+$(1,3), and $\I$ is the identity matrix. Although it is necessary to consider all the structure of $D=10$ Lorentz (Poincar\`e) group,  we can take only the first Casimir operator which defines the invariant ``length"  $p^M p_M$\cite{gamma,RAHA,vio}. So, borrowing the $D=4$ classes of this invariant in order to analyze the spectrum, we can take the massless case \footnote{It is not the only case.}, e.g., 
\begin{eqnarray}
p^M p_M=0 \Rightarrow p^{\mu} p_{\mu}= -p^{a} p_{a}\equiv m^2
\label{pmupmu}
\end{eqnarray} 
where $M=0, \cdots  9$, $\mu=0,1,2,3$ and $a=4,\cdots ,9$. The combination of the moment of the extra dimensions is the mass term of the usual four dimensions. We are going to analyze from the  Dirac matrices in $10D$, what yields the starting space-times $D=1+9$ and $D=5+5$ to arrive to the $D=1+3$ space-time. And $10D$ are the right ones to naturally yield the appearance of $4$ fermionic families in $4D$.

\section{The \texorpdfstring{$D=1+9$}~ Space-time}
%
To set up the Dirac equation in this space-time framework, we need to verify the Lorentz characteristic and the suitable representation of each one of the $\Gamma$ matrices \cite{gamma,RAHA}. In the $D=10$ Minkowskian spacetime dimensions, and with the metric given by ${\rm diag}(+---;------)$. To this aim we have to verify the suitable representation of the Dirac matrices in $D=1+3$ in such way that this set of matrices have to obey the simple Clifford relation $\{\gamma^\mu, \gamma^\nu \} = 2\eta^{\mu\nu}\I$ and that represents a Lorentz group SO$^+$(1,3) and with the metrics $\eta^{\mu\nu} = {\rm diag}(+---)$. Therefore taking them in the Majorana representation, we can write them as
\begin{eqnarray}
\gamma^0=\left(\begin{array}{cc}
0 & \sigma_y \\ \sigma_y & 0
\end{array}\right)~;~~~~
\gamma^1=\left(\begin{array}{cc}
i\sigma_z & 0 \\ 0 & i\sigma_z
\end{array}\right)~;~~~~
\gamma^2=\left(\begin{array}{cc}
0 & -\sigma_y \\ \sigma_y & 0
\end{array}\right)~;~\nonumber\\
\gamma^3=\left(\begin{array}{cc}
-i\sigma_x & 0 \\ 0 &-i\sigma_x
\end{array}\right)~;~~~~
\gamma_5=\left(\begin{array}{cc}
\sigma_y & 0 \\ 0 & -\sigma_y 
\end{array}\right)~;~~~~
\I=\left(\begin{array}{cc}
\I_2&0 \\ 0&\I_2
\end{array}\right)~.~\nonumber
\end{eqnarray}
On the other hand, to the group SO$^+$(1,9), the metric is $\eta^{MN} = {\rm diag}(+---;------)$, and the its Clifford relation can be written as 
\begin{equation}
\{\Gamma^M, \Gamma^N \} = 2\eta^{MN}\I_{10}
\label{diracalg10}
\end{equation}
where the index is now $M=(0, ..., 9)$ and labeled the tensor representation of SO$^+$(1,9) and $\I_{10}$ is the identity matrix of this space. Bearing in mind an {\it{ad hoc}} dimension reduction, it is interesting to separate them in three sets of Dirac matrices of SO$^+$(1,9), where one of the sets will be the target Lorentz group SO$^+$(1,3), containing the usual $\gamma^\mu$,  identity $\I$ and $\gamma_5$ matrices of the $D=1+3$ space-time. So, the Dirac matrices in the Majorana-Weyl representation of the group SO$^+$(1,9), can be shown as
\begin{eqnarray}
\Gamma^\mu&=&\sigma_x\otimes(\gamma^1\gamma^0)\otimes\gamma^\mu~, \hspace{6cm}\mu=0,1,2,3 \\
\Gamma^j&=&\sigma_x\otimes(i\gamma^3\gamma^0, i\gamma^0\gamma^2)\otimes\I;~ i\sigma_y\otimes\gamma_5\otimes\I~, \hspace{2.6cm}j=5,6,7\\
\Gamma^I&=&i\sigma_y\otimes(-i\gamma^1\gamma_5, -\gamma^1)\otimes(i\gamma_5);~ \sigma_x\otimes(-i\gamma^2\gamma^3)\otimes(i\gamma_5)~,~~ I=8,9,4
\end{eqnarray}
On the other hand, the matrices $\sigma_x$ and $i\sigma_y$, which are in front of the matrices definition, split the matrices of the group SO$^+$(1,9) in two blocks with distinct chiralities ($\Sigma^M$ and $\widetilde\Sigma^M$) due to the ``global'' form of the matrices, or
\begin{equation}
  \Gamma^M=\left(\begin{array}{cc}
0&\Sigma^M\\\widetilde\Sigma^M&0
\end{array}\right)
\end{equation}
We can observe from the above $\Gamma^M$ matrices that
\begin{eqnarray}
\widetilde\Sigma^M&=&\Sigma^M ~~ {\rm for} ~~ M=0,1,...,6,~~ \text{due to}~ \sigma_x  \\
\widetilde\Sigma^M&=&-\Sigma^M ~~ {\rm for} ~~ M=7,8,9,~~ \text{due to}~ i\sigma_y 
\end{eqnarray}
The $\Sigma$ matrices can be read off as
\begin{eqnarray}
\Sigma^\mu&=&(\gamma^1\gamma^0)\otimes\gamma^\mu~,\hspace{6cm}~\mu=0,1,2,3 \\
\Sigma^j&=&(i\gamma^3\gamma^0, i\gamma^0\gamma^2, \gamma_5)\otimes\I~,\hspace{4.6cm}j=5,6,7\\
\Sigma^I&=&(-i\gamma^1\gamma_5, -\gamma^1, -i\gamma^2\gamma^3)\otimes(i\gamma_5)~,\hspace{3.1cm}I=8,9,4
\end{eqnarray}
Therefore, the contraction $\Sigma^\mu\partial_\mu$ can be written as
\begin{equation}
  \Sigma^\mu\partial_\mu=(\gamma^1\gamma^0)\otimes\Sl\partial
\end{equation}
where we can identify two different sectors: one concerning to the spinor flavors (F) and one to the Lorentz indexes (L). The matrices which pertain to the ``chiral'' sector $\Sigma$ of the group SO$^+$(1,9) are of the
type F $\otimes$ L  and they are  $4N \times 4N$ matrices, where $N$ is the number of flavors (dimension of F). In the case treated we have $N=4$. The ``chiral'' sector $\widetilde\Sigma$, is furnished with four flavors too, thus the group SO$^+$(1,9) contain a total of eight flavors.

\subsection{The Spinor Fields}

A particular feature of the Weyl, and Majorana-Weyl representations is that they let accommodated the chiral components ($\xi$ and $\chi$) of the spinor field in a simpler way\cite{gamma,RAHA}. Indeed in these representations, the SO$^+$(1,9) spinor field, $X$ and its adjoint  $\overline{X}$ can be read as:
\begin{equation}
X=\left(\begin{array}{c}
\xi   \\   \chi
\end{array}\right)~;~~~~
\overline{X}=X^\dag\Gamma^0=\left(\begin{array}{cc}
\chi^\dag\widetilde\Sigma^0~ & ~\xi^\dag\Sigma^0
\end{array}\right)
\end{equation}
We can emphasize that in the products of the type $\Gamma^M{X}$, the adjoint part, relating to the  $\Sigma^M$, is the lower chiral component  $\chi$, which can be explicit represented showing its components as  $\chi_{a\,\alpha}$, where $a= 1,...,4$ labels the flavor sector and $\alpha=1,...,4$ labels the Lorentz sector. Expanding only in the flavor sector, we have
\begin{equation}
\chi=\left(\begin{array}{c}
\chi_1 \\ \vdots \\ \chi_4
\end{array}\right)
\label{chi}
\end{equation}
In view of a future dimensional reduction and to diagonalize the matrices, it is convenient to introduce the extended adjoint of the target space SO$^+$(1,3), that is defined as 
\begin{equation}
\widetilde\chi=\chi^\dag(\I\otimes\gamma^0)=
\left(\begin{array}{ccc}
\chi_1^\dag&\cdots\chi_4^\dag
\end{array}\right)
\left(\begin{array}{lcr}
\gamma^0~&~&~\\
~&\ddots&~\\
~&~&~\gamma^0
\end{array}\right)=
\left(\begin{array}{ccc}
\overline\chi_1&\cdots\overline\chi_4
\end{array}\right)
\label{chiadj}
\end{equation}
On the other hand, the adjoint of SO$^+$(1,9) is given by
\begin{equation}
   \chi^\dag\widetilde\Sigma^0=\chi^\dag(\gamma^1\gamma^0\otimes\gamma^0)   \label{adjSO1-9}
\end{equation}
We can express $\chi^\dag\widetilde\Sigma^0$ in terms of $\widetilde\chi$ as
\begin{equation}  
\chi^\dag\widetilde\Sigma^0=\widetilde\chi(\gamma^1\gamma^0\otimes\I)
\end{equation}
We now have conditions to construct the Lagrangeans concerning to this space-time.

\subsection{The \texorpdfstring{$D=1+9$}~ Spinor Lagrangean}

Considering that $X=X(x^M)$, the starting free and massless Dirac Langragean density is
\begin{equation}
{\cal L}=\overline{X}\,i\,\Gamma^M\partial_M{X}.
\label{lagDiracfree}
\end{equation}
Then, concerning to the group SO$^+$(1,9), the Lagrangean density can be expressed in term of its chiral components as 
\begin{equation}
     {\cal L}=\chi^\dag\widetilde\Sigma^0\,i\,\Sigma^M\partial_M\chi+\xi^\dag\Sigma^0\,i\,\widetilde\Sigma^M\partial_M\xi
\end{equation}
We are keeping, for the sake of simplicity, the same symbol ${\cal L}$ though it only contains the chiral component $\chi$. For the dimensional reduction, we are going to isolate the Lorentz sector SO$^+$(1,3) labeled by the index $\mu$, and for the remaining part of ${\cal L}$ we are going to consider the index $j=5,6,7$ that label the set of matrices proportional to the identity $\I$, and  $I=8,9,4$ that label the set of matrices proportional to  $\gamma_5$. So we can span the coordinates as $x^M=(x^\mu; y^j; Y^I)$ in such way that $\partial_{\mu}x_\nu=\eta_{\mu\nu}$, $\partial_iy_j=\eta_{ij}$ and $\partial_IY_J=\eta_{IJ}$. Also, for the sake of simplicity, we are going to refer $x^\mu$ simply as $x$, so the application of the differential operator  $i\Sigma^M\partial_M$  to a plane wave solution of the type $\chi(x)e^{[-ip^jy_j-ip^JY_J]}$ is identical the application the operator:
\begin{equation}
i\,\Sigma^M\partial_M=i\,\Sigma^\mu\partial_\mu+\Sigma^jp_j+\Sigma^Ip_I   \label{slash4}
\end{equation}
Which we can explicitly shown as
\begin{eqnarray}
i\,\Sigma^\mu\partial_\mu&=& (\gamma^1\gamma^0)\otimes(i\Sl\partial) \label{slash1}\\
\Sigma^ip_i         &=& -r\otimes\I\\
\Sigma^Ip_I         &=& -s\otimes(i\gamma_5)
\end{eqnarray}
where $r=-(i\gamma^3\gamma^0p_5+i\gamma^0\gamma^2p_6+\gamma_5p_7)$ and $s=-(-i\gamma^1\gamma_5p_8-\gamma^1p_9-i\gamma^2\gamma^3p_4)$. Then we can write down the \eqref{slash4} as
$
i\,\Sigma^M\partial_M=\gamma^1\gamma^0\otimes(i\Sl\partial)-r\otimes\I-s\otimes(i\gamma_5).
$
So the part of the Lagrangean density that contains $\chi(x)$ and its adjoint partner $\widetilde\chi$ is written as 
\begin{equation}
     {\cal L}=\widetilde\chi(\gamma^1\gamma^0\otimes\I)\,[\gamma^1\gamma^0\otimes{i\Sl\partial}-r\otimes\I-s\otimes(i\gamma_5)]\chi .
\end{equation}
In order to compute the spectrum of the model we can introduce the matrices $A=\gamma^1\gamma^0\,r$ and $B=\gamma^1\gamma^0\,s$ which yields
\begin{equation}
     {\cal L}=\widetilde\chi\,[\I\otimes{i\Sl\partial}-A\otimes\I-B\otimes(i\gamma_5)]\chi ,
\end{equation}
where the matrices $A$ and $B$ have the form:
\begin{equation}
A=\left( \begin{array}{cccc} 
0&ip_5&ip_7&ip_6 \\ -ip_5&0&ip_6&-ip_7 \\ -ip_7&-ip_6&0&ip_5 \\ -ip_6&ip_7&-ip_5&0
\end{array}\right)~;~~~~~
B=\left(\begin{array}{cccc}
0&ip_4&-ip_8&-ip_9\\ -ip_4&0&ip_9&-ip_8\\ ip_8&-ip_9&0&-ip_4\\ ip_9&ip_8&ip_4&0
\end{array}\right)
\end{equation}
which are both Hermitian. We emphasize that they commute each other, therefore using the unitary matrix $U$ that simultaneously diagonalize $A$ and $B$, we obtain real eigenvalues. So from $A$ we obtain that  
\begin{equation}
\{\varphi,\varphi,-\varphi,-\varphi\}; ~~~\varphi\equiv\sqrt{p_5^2+p_6^2+p_7^2}.
\end{equation}
And the eigenvalues of $B$ are
\begin{equation}
\{\varphi_5,\varphi_5,-\varphi_5,-\varphi_5\}; ~~~\varphi_5\equiv\sqrt{p_8^2+p_9^2+p_4^2}
\end{equation}
Considering
\begin{equation}
  \chi=(U\otimes\I)\psi \;\;\; \mbox{and} \;\;\; \widetilde\chi=\widetilde\psi(U^{\dag}\otimes\I),
\end{equation}
then, the Lagrangean density ${\cal L}$, can be expressed in terms of the components of $\psi$ as 
\begin{eqnarray}
  {\cal L}&=&\overline\psi_1(i\Sl\partial+\varphi-i\gamma_5\varphi_5)\psi_1+\overline\psi_2(i\Sl\partial-\varphi-i\gamma_5\varphi_5)\psi_2\nonumber\\&+&\overline\psi_3(i\Sl\partial+\varphi+i\gamma_5\varphi_5)\psi_3+\overline\psi_4(i\Sl\partial-\varphi+i\gamma_5\varphi_5)\psi_4.
\end{eqnarray}
So, we can observe that, after a dimensional reduction, we arrive at an expression that revealed terms of the usual Yukawa type and of a pseudo-Yukawa type, which appear as a by-pass result of the reduction of the $6$ extra dimensions.
Although these terms we have called Yukawa, actually they can be interpreted as condensation of scalar fields which have dimension of mass and a status of mass term. Furthermore, from the analysis of the discrete symmetries applied to these terms we verify (see Appendix) that what we have called usual Yukawa term is also invariant under {\em CP} symmetry and the pseudo-Yukawa term violates the {\em CP} symmetry. We remark that the last feature indicates that this model could accommodate the dynamics of neutrinos. 

\section{The \texorpdfstring{$D=5+5$}~ Space-time}

Another signature of the $D=10$ space-time is concerned to one that contains five space and five time directions, or $D=5+5$, endowed with a pseudo-Euclidean metric. This signature is not arbitrary. Indeed it comes from supersymmetry, where in particular it is one of the possible signatures as a result of Majorana-Weyl constraint applied to superstrings models which implies that the only allowed $D=10$ space-time signatures are $D=1+9$, $D=9+1$ and $D=5+5$. The first two are the usual ones, and their consequences to the superstrings were exhaustively explored in the literature. The third one is discarded due to the {\it a priori} consequent problems as inconsistencies of the moment definition, and the dynamics of the compactified space-time where it generates ghosts. 
In spite these {\it a priori} problems, we would like to show that it deserves a more deep analysis before a simply discarding. In fact, its own representation reveal an intriguing two dimensional block (of dimensions) characteristics that plays an interesting role in the dimension reduction.   On the other hand and unfortunately, superstrings has no answer to the question why our universe is (at least apparently) in four dimensions, neither why it has the signature $D=1+3$ \cite{duff}. So we use the {\ it ad hoc} argument to reach the usual four dimensional and reaching it using a background Lorentz symmetry as it basis.  

The representation of the  Dirac $\Gamma $ matrices to the space-time $D=5+5$ can be written in an iterative way from their $4D$ ones \cite{gamma},
\begin{eqnarray}
\Gamma ^{M}=(\Gamma ^{m},\Gamma ^{\overline{m}})\Longrightarrow \left\{ 
\begin{array}{l}
\Gamma^{m}=~\sigma_x \otimes\,\I\,\otimes \gamma ^{m} \\ 
\Gamma ^{\overline{m}}=i \sigma_y \otimes \gamma ^{m}\otimes\I
\end{array}
\right. \,,
\label{Gamma5+5}
\end{eqnarray}
where $M=0,...,9$, $m=0,...,4$, $\overline{m}=m+5$. We have two possibilities of the $\gamma^{4}$ matrix which give us two different signatures. In the case where we take $\gamma ^{4}=\gamma_{5}$ we have the signature $(+---+;-+++-)$. On the other hand, taking $\gamma ^{4}=i\gamma _{5}$, the signature becomes $(+----;-++++)$ , with $\gamma _{5}=i\gamma ^{0}\gamma ^{1}\gamma ^{2}\gamma ^{3}$. It is interesting to analyze the spectra of these two signatures.

It is remarkable that $D=5+5$ provides only two sets of signatures which lead to different physical spectra after dimension reduction, and they have two signs as the only difference\cite{gamma}. For instance, considering the massless case of the mass-energy relation on the full $10D$ space, we can reach the only two target signature spaces-times signed above. The first one, we take ($\gamma ^{4}=\gamma _{5}$) where we use the first choice for the space-time signature target space-time  \cite{dirac,gamma,vio} has the invariant: 
\begin{eqnarray}
p^{M}p_{M}
&=&p_{0}^{2}-p_{1}^{2}-p_{2}^{2}-p_{3}^{2}+p_{4}^{2}-p_{5}^{2}+p_{6}^{2}+p_{7}^{2}+p_{8}^{2}-p_{9}^{2}=0
\label{pp23}
\end{eqnarray}
The second choice for the space-time signature ($\gamma ^{4}=i\gamma _{5}$), as a target space-time, we have the invariant: 
\begin{eqnarray}
p^{M}p_{M}
&=&p_{0}^{2}-p_{1}^{2}-p_{2}^{2}-p_{3}^{2}-p_{4}^{2}-p_{5}^{2}+p_{6}^{2}+p_{7}^{2}+p_{8}^{2}+p_{9}^{2}=0
\label{pp55}
\end{eqnarray}
In these two cases we are interested to obtain the spectra and their features to verify the possibility to include See-saw mechanism, {\em CP} violation, and a fourth-generation neutrino\cite{RAHA}.

Analogous to the previous case, here the Dirac matrices also obey the algebra \eqref{diracalg}. Nevertheless the Dirac matrices in the chiral representation of the Lorentz group  SO$^+$(1,3) are now,
\begin{eqnarray}
\gamma^0=\left(\begin{array}{cc}
0 & -\I_2 \\ -\I_2 & 0
\end{array}\right)~;~~~~
\gamma^1=\left(\begin{array}{cc}
0 & \sigma_x \\ -\sigma_x & 0
\end{array}\right)~;~~~~
\gamma^2=\left(\begin{array}{cc}
0 & \sigma_y \\ -\sigma_y & 0
\end{array}\right)~;~\nonumber\\
\gamma^3=\left(\begin{array}{cc}
0 & \sigma_z \\ -\sigma_z & 0
\end{array}\right)~;~~~~
\gamma_5=\left(\begin{array}{cc}
\I_2 & 0 \\ 0 & -\I_2
\end{array}\right)~;~~~~
\I=\left(\begin{array}{cc}
\I_2&0 \\ 0&\I_2
\end{array}\right)~.~\nonumber
\label{Diracmat1+4}
\end{eqnarray}
We are going to analyze the dimension reductions passing through two different intermediate space-time signatures.

\subsection{The Intermediate \texorpdfstring{$D=2+3$}~ Space-time}

As it was emphasized in the beginning of the Section, the difference of the previous case is to take $\gamma^4=\gamma_5$, and so all the definitions and computations are analogous, therefore the $\Gamma$ matrices in the Weyl representation of the SO$^+$(5,5) group, can be represented as 
\begin{eqnarray}
\Gamma^\mu&=&\sigma_x\otimes\I\otimes\gamma^\mu, \hspace{4,0cm}\mu=0,1,2,3 \label{55iiGammas1}\\
\Gamma^I  &=&\sigma_x\otimes\I\otimes\gamma_5,~~ \hspace{3,7cm}I=4\label{55iiGammas2}\\
\Gamma^j  &=&i\sigma_y\otimes(\gamma^0, \gamma^1, \gamma^2, \gamma^3, \gamma_5)\otimes\I \hspace{1.4cm}j=5,6,7,8,9 
\label{55iiGammas3}
\end{eqnarray} 
Again the matrices $\sigma_x$ and $i\sigma_y$ in front of the products in the above equations separate the matrices in two blocks of distinct chiralities, or
\begin{equation}
  \Gamma^M=\left(\begin{array}{cc}
0&\Sigma^M\\\widetilde\Sigma^M&0
\end{array}\right)
\end{equation}
where $\Sigma^M$ and $\widetilde\Sigma^M$ represent these chiralities. So for the above matrices we can say that 
\begin{eqnarray}
\widetilde\Sigma^M&=&\Sigma^M ~~ {\rm for} ~~ M=0,1,2,3,4~~ \text{due to}~ \sigma_x  \\
\widetilde\Sigma^M&=&-\Sigma^M ~~ {\rm for} ~~ M=5,6,7,8,9~~ \text{due to}~ i\sigma_y 
\end{eqnarray}
and the $\Sigma$ matrices are
\begin{eqnarray}
\Sigma^\mu&=&\I\otimes\gamma^\mu\;,~~\hspace{4cm} \mu=0,1,2,3 \\
\Sigma^I&=&\I\otimes\gamma_5\;,~~\hspace{4cm}I=4 \\
\Sigma^j&=&(\gamma^0, \gamma^1, \gamma^2, \gamma^3, \gamma_5)\otimes\I\;,~~\hspace{1.4cm} j=5,6,7,8,9 .
\end{eqnarray}
Thus the contraction $\Sigma^\mu\partial_\mu$ can be expressed in the way 
\begin{equation}
   \Sigma^\mu\partial_\mu=\I\otimes\Sl\partial .
\end{equation}
In these representation, the spinor $X$ and its adjoint $\overline{X}$ fields in SO$^+$(5,5) can be read as
\begin{equation}
X=\left(\begin{array}{c}
\xi   \\   \chi
\end{array}\right)~;~~~~
\overline{X}=X^\dag\Gamma^0\Gamma^4\Gamma^6\Gamma^7\Gamma^8=\left(\begin{array}{cc}
\chi^\dag{\cal A}~ & ~-\xi^\dag{\cal A}
\end{array}\right)
\end{equation}
where
\begin{equation}
{\cal A}=\Sigma^0\Sigma^4\Sigma^6\Sigma^7\Sigma^8 = (-i\gamma^0\gamma_5)\otimes(\gamma^0\gamma_5) .
\end{equation}
And so, by analogous to the previous case and straightforward computation we obtain the Lagrangean,  
\begin{equation}
     {\cal L}=-\widetilde\chi(i\gamma^0\gamma_5\otimes\gamma_5)[\I{\otimes}i{\Sl\partial}-r\otimes\I-s\otimes(i\gamma_5)]\chi ,
\end{equation}
where $r=- (\gamma^0p_5+\gamma^1p_6+\gamma^2p_7+\gamma^3p_8+\gamma_5p_9$ and $s=ip_4$. In terms of the new matrices $K=-i\gamma^0\gamma_5$, $A=-i\gamma^0\gamma_5\,r$ and $B=-i\gamma^0\gamma_5\,s$, we have 
\begin{equation}
     {\cal L}=\widetilde\chi(\I\otimes\gamma_5)[K\otimes\,{i\Sl\partial}-A\otimes\I-B\otimes(i\gamma_5)]\chi.
\end{equation}
To diagonalize ${\cal L}$, we propose, by means of a matrix $\Omega$, the transformations of the component spinor and its adjoint, respectively as 
\begin{equation}
  \chi=(\Omega\otimes\I)\psi \;\;\; \mbox{and} \;\;\; \widetilde\chi=\widetilde\psi(\Omega^{\dag}\otimes\I).
\end{equation}
In terms of $\psi$ and $\widetilde\psi$, the Lagrangean density can be re-written as,
\begin{equation}
    {\cal L}=\widetilde\psi(\Omega^{\dag}\otimes\I)(\I\otimes\gamma_5)\left[K{\otimes}{i\Sl\partial}-A{\otimes}\I-B{\otimes}(i\gamma_5)\right]({\Omega}\otimes\I)\psi.
\end{equation}
The proposed matrix $\Omega$, which diagonalize the matrix components of the above Lagrangean, leads to diagonal matrices $\ell$ and $\ell_5$  in such way that,
\begin{eqnarray}
{\Omega}^{\dag}K{\Omega}&=&\gamma_5 ,\nonumber \\
{\Omega}^{\dag}A{\Omega}&=&\gamma_5\ell, \\
{\Omega}^{\dag}B{\Omega}&=&\gamma_5\ell_5. \nonumber
\label{diag23}
\end{eqnarray}
What permit us to obtain the properly diagonalized Lagrangean:
\begin{equation}
    {\cal L}=\widetilde\psi(\gamma_5\otimes\gamma_5 )\left[\I{\otimes}{i\Sl\partial}-\ell{\otimes}\I-\ell_5{\otimes}(i\gamma_5)\right]\psi .
\end{equation}
The relations to the diagonal matrices \eqref{diag23} given above can be re-written as
\begin{eqnarray}
&&{\Omega}\gamma_5{\Omega}^{\dag}=K,  \nonumber \\  
&&{\Omega}^{-1}K\!A\,{\Omega}=\ell, \label{OKAOii}\\
&&K\!B=\ell_5 . \nonumber
\end{eqnarray}
From these equations, we can read the elements of the diagonal matrices $\ell$ and $\ell_5$, respectively, as the eigenvectors of $K\!A$ and $K\!B$ matrices. We emphasize that the matrices $A$ and $B$ are both Hermitian, but they do not commute with each other. On the other hand, $K\!A=r$ is not Hermitian and $K\!B=s$ is proportional to the identity and is real, so $K\!A$ commutes with $K\!B$, so
\begin{equation}
K\!A=\left( \begin{array}{cccc} 
-p_9&0&p_5-p_8&-p_6+ip_7\\ 0&-p_9&-p_6-ip_7&p_5+p_8\\ p_5+p_8&p_6-ip_7&p_9&0 \\ p_6+ip_7&p_5-p_8&0&p_9
\end{array}\right)~;~~
K\!B=ip_4\I
\label{55iiKAeKB}
\end{equation}
So, $\pm\sqrt{p_5^2-p_6^2-p_7^2-p_8^2+p_9^2}=\pm\sqrt{-p^ip_i}$ are the eigenvalues of $K\!A$. If we allow $m$ to satisfy the mass constraint in SO$^+$(1,3) given by $p^{\mu}p_\mu=m^2$, it results that $m$ is positive, and from Eq.~(\ref{pp23}) it follows that $m^2+p_4^2+p^ip_i=0$. Thus the eigenvalues of $K\!A$ can be read as $\pm\sqrt{-p_4^2+m^2}$. The Lagrangean density ${\cal L}$ can be expanded in terms of its component 
SO$^+$(1,3) spinors fields as
\begin{eqnarray}
  {\cal L}
&=& \overline\psi_1\left(\gamma_5\,i\Sl\partial-{\tsty\sqrt{p_4^2+m^2}}~\I-p_4\gamma_5\right)\!\psi_1
	 +\overline\psi_2\left(\gamma_5\,i\Sl\partial-{\tsty\sqrt{p_4^2+m^2}}~\I-p_4\gamma_5\right)\!\psi_2\nonumber\\
&-& \overline\psi_3\left(\gamma_5\,i\Sl\partial+{\tsty\sqrt{p_4^2+m^2}}~\I-p_4\gamma_5\right)\!\psi_3
   -\overline\psi_4\left(\gamma_5\,i\Sl\partial+{\tsty\sqrt{p_4^2+m^2}}~\I-p_4\gamma_5\right)\!\psi_4\nonumber
	\label{55iiexpand}\\
\end{eqnarray} 
So the eigenvalues of the matrix of mass can be directly read from the expanded Lagrangean density \eqref{55iiexpand}, namely,
\begin{eqnarray}
m_{1} &=&  p_4+\sqrt{p_4^{2}+m^{2}},\nonumber\\
m_{2} &=& -p_4+\sqrt{p_4^{2}+m^{2}}, \label{ss1}\\
m_{3} &=&  p_4-\sqrt{p_4^{2}+m^{2}},\nonumber \\
m_{4} &=& -p_4-\sqrt{p_4^{2}+m^{2}}. \nonumber
\end{eqnarray}
each one of the above eigenvalues is degenerated four times. Therefore this target model could fit a See-saw type mechanism to the neutrino masses. To this aim it is useful to illustrate our discussion presenting a Table \ref{table1} with the eigenvalues of the matrix of mass, \eqref{ss1}, and some particular values of $p_4$.
\begin{table}[h]
  \centering
  \large
  \begin{eqnarray}
  \begin{array}{|c|c|c|c|} \hline
  & p_4 \rightarrow 0 &p_4 \rightarrow \infty & p_4 \gg m\\ \hline
 m_1 &  m & \infty  & 2p_4\\ \hline
 m_2&   m & 0       & \frac{m^2}{2p_4}\\ \hline
 m_3 & -m & 0       &-\frac{m^2}{2p_4}\\ \hline
 m_4 & -m & -\infty &-2p_4\\\hline
  \end{array}\nonumber
  \end{eqnarray}
  \caption{Eigenvalues of the mass matrix with different choices for $p_4$.}
  \label{table1}
\end{table} 
%
We notice that whenever $p_4 \rightarrow 0$, we get degenerate massive Dirac equations. In the case that $p_4\rightarrow \infty$ generates masses of two types: one almost zero and a very massive one in $D=1+3$.
So we particularly remark that, in Table \ref{table1}, the case where $p_4 \gg m$, namely
\begin{eqnarray}
m_{1} &\approx& 2p_4,\\
m_{2} &\approx& \frac{m^2}{2p_4}.
\end{eqnarray}
Indeed this case sets up a mechanism similar to the usual See-Saw Mechanism for neutrino masses \cite{RAHA,fil}, so that this proposal could allow us to understand the generation of hierarchies for the fermion masses in $D=1+3$.

Finally, in Figure 1, we present the eigenvalues of the matrix mass as a
function of the momentum $p_4$.

\begin{figure}[h]
\begin{center}
\includegraphics[angle=0,width=0.4\textwidth]
{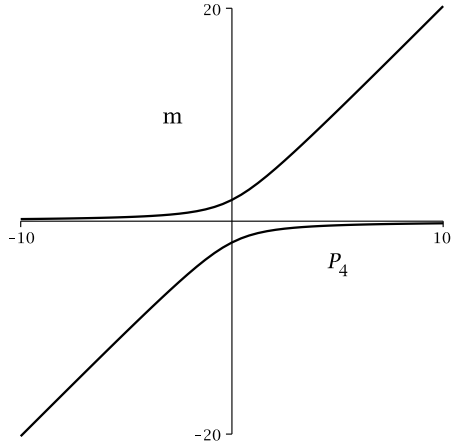}%
\end{center}
\caption{\label{fig1} \it The eigenvalues of the mass matrix in D=1+3 dimensions for a fixed m=4.}
\end{figure}

\subsection{The Intermediate \texorpdfstring{$D=1+4$}~ Space-time}

To the group SO$^+$(5,5) the metric tensor is $\eta^{MN} = {\rm diag}(+----;-++++)$,  and the relation amongst the Dirac matrices of the group  SO$^+$(5,5) and the metric tensor is given by the same relation as the previous case \eqref{diracalg10}, namely 
\begin{equation}
\{\Gamma^M, \Gamma^N \} = 2\eta^{MN}\I_{10}
\end{equation}
where the index $M=0,..., 9$ label the tensor representations of SO$^+$(5,5), and $\I_{10}$ is the identity matrix. Using an analogous to the previous procedure to a future dimension reduction it is interesting to separate in three sets of Dirac matrices of SO$^+$(5,5), particularly, one set related to the Lorentz group SO$^+$(1,3) or related to the $\gamma_{\mu}$, and another related to the $\gamma_5$, and also another related to identity matrix. So, the Dirac matrices of the group SO$^+$(5,5) in the Majorana-Weyl representation are written as
\begin{eqnarray}
\Gamma^\mu&=&\sigma_x\otimes\I\otimes\gamma^\mu, \hspace{4,0cm}\mu=0,1,2,3 ; \nonumber \\
\Gamma^I  &=&\sigma_x\otimes\I\otimes(i\gamma_5),~~ \hspace{3,3cm}I=4;\label{55iGammas2}\\
\Gamma^j  &=&i\sigma_y\otimes(\gamma^0, \gamma^1, \gamma^2, \gamma^3, i\gamma_5)\otimes\I \hspace{1,3cm}j=5,6,7,8,9 .
\nonumber
\end{eqnarray}
Then we can read the $\Sigma$ matrices as 
\begin{eqnarray}
\Sigma^\mu&=&\I\otimes\gamma^\mu\;,~~\hspace{4cm} \mu=0,1,2,3; \nonumber \\
\Sigma^I&=&\I\otimes(i\gamma_5)\;,~~\hspace{3.6cm}I=4 \\
\Sigma^j&=&(\gamma^0, \gamma^1, \gamma^2, \gamma^3, i\gamma_5)\otimes\I\;,~~\hspace{1.3cm} j=5,6,7,8,9 . \nonumber
\end{eqnarray}
In this space-time due to the extra time dimensions the spinor $X$ and its the adjoint spinor $\overline{X}$ concerning the group  SO$^+$(5,5) are represented as 
\begin{equation}
X=\left(\begin{array}{c}
\xi   \\   \chi
\end{array}\right)~;~~~~
\overline{X}=X^\dag\Gamma^0\Gamma^6\Gamma^7\Gamma^8\Gamma^9=\left(\begin{array}{cc}
\chi^\dag{\cal A}~ & ~\xi^\dag{\cal A}
\end{array}\right), 
\end{equation}
where ${\cal A}=\Sigma^0\Sigma^6\Sigma^7\Sigma^8\Sigma^9= \gamma^0\otimes\gamma^0$. 
In order to do a dimension reduction and to diagonalize the matrix components, we use the same as the previous procedure. And analogous to the  Eq. \eqref{chiadj} of the adjoint concerning the group SO$^+$(1,3), the adjoint spinor related to the group SO$^+$(5,5) is given by
\begin{equation}
   \chi^\dag{\cal A}=\chi^\dag(\gamma^0\otimes\gamma^0),   \label{adjSO55i}
\end{equation}
then, we can express $\chi^\dag{\cal A}$ in terms of $\widetilde\chi$ as
\begin{equation}   
\chi^\dag{\cal A}=\widetilde\chi(\gamma^0\otimes\I).
\end{equation}
Considering that $X=X(x^M)$ to built the Lagrangean density we start from a free and massless Dirac one \eqref{lagDiracfree} which obeys the mass constraint \eqref{pp55}, and, concerning to SO$^+$(5,5), it can be expressed in terms of chiral component fields as
\begin{equation}
     {\cal L}=\chi^\dag{\cal A}\,i\,\Sigma^M\partial_M\chi+\xi^\dag{\cal A}\,i\,\widetilde\Sigma^M\partial_M\xi .
\end{equation}
As in previous cases, we take only the part of the Lagrangean density ${\cal L}$ that contains the chiral component field $\chi$, but for simplicity we maintain the same symbol. In this case, to the dimension reduction procedure we will separate the Lorentz sector SO$^+$(1,3) that is labeled by the index $\mu$, the index $I=4$ that labels the part proportional to $\gamma_5$, and the index $j=5,6,7,8,9$ that labels the part proportional to $\I$. So we can represent the coordinates as  $x^M=(x^\mu; Y^I; y^j)$ in such way that  $\partial_{\mu}x_\nu=\eta_{\mu\nu}$, $\partial_IY_J=\eta_{IJ}$ and $\partial_iy_j=\eta_{ij}$. And again, for the sake of simplicity, we are ignoring the index of $x$ the plane wave solution can be written as $\chi(x)e^{(-ip^JY_J-ip^jy_j)}$, and the application of the differential operator $i\Sigma^M\partial_M$ is, analogous to the previous cases, so
\begin{equation}
i\,\Sigma^M\partial_M=i\,\Sigma^\mu\partial_\mu+\Sigma^ip_i+\Sigma^Ip_I  
\label{sigmaD55}
\end{equation}
where the $\Sigma$ matrices can be written, in this case, as
\begin{eqnarray}
i\,\Sigma^\mu\partial_\mu&=& \I\otimes{i\Sl\partial} \nonumber\\
\Sigma^jp_j         &=& -r\otimes\I\\
\Sigma^Ip_I         &=& -s\otimes(i\gamma_5), \nonumber
\end{eqnarray}
where $r=-(\gamma^0p_5+\gamma^1p_6+\gamma^2p_7+\gamma^3p_8+i\gamma_5p_9)$ and $s=-p_4\I$. By means of the last definitions the operator \eqref{sigmaD55} can be written as
\begin{equation}
i\,\Sigma^M\partial_M=\I\otimes{i\Sl\partial}-r\otimes\I-s\otimes(i\gamma_5).
\end{equation}
Therefore the part of the Lagrangean that contains $\chi(x)$ and $\widetilde\chi$ can be obtain as
\begin{equation}
     {\cal L}=\widetilde\chi(\gamma^0\otimes\I)\,[\I\otimes{i\Sl\partial}-r\otimes\I-s\otimes(i\gamma_5)]\chi
\end{equation}
So, in term of the new matrices $K=\gamma^0$, $A=\gamma^0\,r$ and $B=\gamma^0\,s$ we can write them as
\begin{equation}
     {\cal L}=\widetilde\chi\,[K\otimes{i\Sl\partial}-A\otimes\I-B\otimes(i\gamma_5)]\chi
\end{equation}
To diagonalize the ${\cal L}$, we propose, by means of a matrix $\Omega$, the transformations of the component spinor and its adjoint, respectively as 
\begin{equation}
  \chi=(\Omega\otimes\I)\psi \;\;\; \mbox{and} \;\;\; \widetilde\chi=\widetilde\psi(\Omega^{\dag}\otimes\I).
\end{equation}
In terms of $\psi$ and $\widetilde\psi$, the Lagrangean density can be re-written as
\begin{equation}
    {\cal L}=\widetilde\psi(\Omega^{\dag}\otimes\I)\left[K{\otimes}{i\Sl\partial}-A{\otimes}\I-B{\otimes}(i\gamma_5)\right]({\Omega}\otimes\I)\psi.
\end{equation}
We have proposed that the matrix $\Omega$ which diagonalize the component matrices of the above Lagrangean can lead to diagonal matrices $\ell$ and $\ell_5$. Very similar to the previous case, we are leaded to the same set o equations which permit us to obtain the properly diagonalized Lagrangean as
\begin{equation}
    {\cal L}=\widetilde\psi(\gamma_5\otimes\I )\left[\I{\otimes}{i\Sl\partial}-\ell{\otimes}\I-\ell_5{\otimes}(i\gamma_5)\right]\psi .
\end{equation}
In analogue way, we can read the elements of the diagonal matrices $\ell$ and $\ell_5$ respectively as the eigenvectors of $K\!A$ and $K\!B$ matrices, and we can observe that the matrices $A$ and $B$ are both Hermitian, but they do not commute with each other. Again, $K\!A=r$ is not Hermitian and $K\!B=s$ is proportional to the identity and real, so $K\!A$ commutes with $K\!B$, namely, in this case,
\begin{equation}
K\!A=\left( \begin{array}{cccc} 
-ip_9&0&p_5-p_8&-p_6+ip_7\\ 0&-ip_9&-p_6-ip_7& p_5+p_8\\ p_5+p_8&p_6-ip_7&ip_9&0 \\ p_6+ip_7&p_5-p_8&0&ip_9
\end{array}\right)~;~~~~~
K\!B=-p_4\I \label{55iAeB}
\end{equation}
In this case, the eigenvalues of $K\!A$ are $\pm\sqrt{p_5^2-p_6^2-p_7^2-p_8^2-p_9^2}=\pm\sqrt{-p^ip_i}\,$. If $m$ satisfy the mass constraint required in SO$^+$(1,3), given by  $p^{\mu}p_\mu=m^2$, we can remark that $m$ is positive, and from Eq.~(\ref{pp55}) it follows that $m^2+p_4^2+p^ip_i=0$. Then the eigenvalues of $K\!A$ can be read as $\pm\sqrt{-p_4^2+m^2}$. So, in analogous computation as the previous section, we arrive to the Lagrangean density in term of the SO$^+$(1,3) spinor fields as
\begin{eqnarray}
  {\cal L}
&=& \overline\psi_1\left(i\Sl\partial-{\tsty\sqrt{-p_4^2+m^2}}~\I+ip_4\gamma_5\right)\!\psi_1
	 +\overline\psi_2\left(i\Sl\partial-{\tsty\sqrt{-p_4^2+m^2}}~\I+ip_4\gamma_5\right)\!\psi_2\nonumber\\
&-& \overline\psi_3\left(i\Sl\partial+{\tsty\sqrt{-p_4^2+m^2}}~\I+ip_4\gamma_5\right)\!\psi_3
   -\overline\psi_4\left(i\Sl\partial+{\tsty\sqrt{-p_4^2+m^2}}~\I+ip_4\gamma_5\right)\!\psi_4\nonumber
	\label{55iexpand}\\
\end{eqnarray}
So the eigenvalues of the mass matrix can be directly read off from the previous expanded Lagrangean density, or 
\begin{eqnarray}
\lambda_1&=&-ip_4+\sqrt{-p_4^2+m^2}, \nonumber \\  
\lambda_2&=& ip_4+\sqrt{-p_4^2+m^2}, \\
\lambda_3&=&-ip_4-\sqrt{-p_4^2+m^2}, \nonumber\\ 
\lambda_4&=& ip_4-\sqrt{-p_4^2+m^2}, \nonumber
\end{eqnarray}
all eigenvalues can be pure-imaginaries ($p_4{\geq}m$), or complex-valued ($p_4<m$) which can lead to {\em CP} violation.

\section{Conclusions}

We have shown that the Dirac equation, if formulated for non-interacting Majorana-Weyl spinors in $D=5+5$ and $D=1+9$, may lead to an interpretation of massive Majorana spinors in $D=1+3$ and in $D=2+3$, after a suitable dimension reduction based on the Lorentz invariance. We confirm the process of reducing the extra dimensions as a mechanism of mass generation. The eigenvectors of the interaction or mass matrix are two particles of opposite chiralities  and different masses moving in $D=1+3$.

The eigenvalues of the mass matrix depend on the energy $p_4$  and $p^4$ (both are associated to chirality due to the $\gamma_5$ matrix) and the invariant four-dimensional space-time mass term $m$. Taking the result for the eigenvalues, we observe, firstly in a free dynamical model and taking $p_4=0$ we obtain four particles with degenerate masses. Second, taking the limit $p_4\rightarrow\infty$, we obtain a massless along with heavy
chiral particles. Finally, when $p_4 \gg m$, we have two Majorana particles of masses $2p_4$ and $\frac{m^{2}}{2p_4}$, respectively. This result is in agreement with the See-Saw Mechanism for neutrinos masses. Therefore, this mechanism of dimension reduction for the generation of Majorana mass can be used to obtain the See-Saw masses for the fermions\cite{RAHA}.

On the other hand, the Majorana-Weyl spinors in Weyl representation of the Dirac matrices in $D=5+5$ shows a very peculiar structure that allows us to naturally separate in two identical blocks of five dimension spacetime with the signatures $D=1+4$ or $D=2+3$ having the mass as the connecting term of the space-times\cite{gamma}. These blocks can reach to a massive four-dimension model controlled by the fifth component (in the moment space) of the block which contain the dynamics. The $D=2+3$ space-time could also be related to the AdS/CFT correspondence\cite{witten} in such way that the connecting term could be interpreted as the cosmological constant.  

In the case $D=1+9$, it is important to note that, the main physical result is the generation of four particles with degenerate masses in $D=1+3$. We can also point out that this proposal allows us to understand the generation of hierarchies for the fermionic masses in $D=1+3$.

Neutrino masses and neutrino oscillations are topics of major relevance in the frame of BSM Physics. Recently, in the work \cite{Ren-Pan}, the authors present a non-trivial connection between the sign of the cosmological constant and the neutrinos oscillation length. Also, Altshuler, in the papers \cite{Altshuler}, presents a very interesting toy-model in which fluxes that exist in extra dimensions may replace the Higgs condensates in vacuum. We believe that the results of our present work point to the possibility that Altshuler raises up that his proposal should be supplemented by the generation of a see-saw scale for the Majorana masses. This opens up a nice perspective to our attempt and we shall be analyzing in more details the way to match our results with the ones in the frame of Altshuler's approach.
 
Finally we would like to speculate about the possibility of these models could to reach to fourth generation of fermions. This question is of special relevance in connection with the discussion of the strongly-coupled electroweak breaking. Indeed if we reduce the model from $10$ to $4$ dimensions, it becomes clear that four neutrinos generations may appear in the $4D$ which remarkably have their genesis in $10$ space-time dimensions. The hierarchy of neutrino masses may also be generated if we suitable fine-tune the parameters in such a way that the fourth generation comes out much heavier than the $e$, $\mu$ and $\tau$- neutrinos.

\appendix
\section{Discrete symmetries}

The transformations related to the discrete symmetries are introduced in the spinor fields as the usual way as the $D=1+3$ space-time, which, up to phases that can be absorbed, are given by 
\begin{eqnarray}
{\cal P}\,\psi(t,{\bf x})\,{\cal P}^\dag&=&\gamma^0\psi(t,-{\bf x}) \, , \\
{\cal C}\,\psi(t,{\bf x})\,{\cal C}^\dag&=&\gamma^0{\overline\psi}^t(t,{\bf x}) \, , \\
{\cal T}\,\psi(t,{\bf x})\,{\cal T}^\dag&=&\gamma^0\gamma_5\psi(-t,{\bf x}) \, , \\
{\cal CP}\,\psi(t,{\bf x})\,({\cal CP})^\dag&=&\I{\overline\psi}^t(t,-{\bf x}) \, , \\
{\cal CPT}\,\psi(t,{\bf x})\,({\cal CPT})^\dag&=&\gamma^0\gamma_5{\overline\psi}^t(-t,-{\bf x}) \, .
\end{eqnarray}
where we can call them respectively {\em P}, {\em C}, {\em T}, {\em CP} and {\em CPT} symmetries. 
Recalling that the usual free massive Dirac Lagrangean is invariant under these discrete transformations, it is clear that the free part of the Dirac Lagrangean is invariant under the above discrete transformations too. So we have to verify this condition to the remaining part of this Lagrangean.

Indeed the Yukawa-type terms contain scalars (fields) which have also standard discrete transformations,
\begin{eqnarray}
{\cal P}\,\varphi(t,{\bf x})\,{\cal P}^\dag&=&\varphi(t,-{\bf x}) \, , \\
{\cal C}\,\varphi(t,{\bf x})\,{\cal C}^\dag&=&{\varphi}^\ast(t,{\bf x}) \, , \\
{\cal T}\,\varphi(t,{\bf x})\,{\cal T}^\dag&=&\varphi(-t,{\bf x}) \, , \\
{\cal CP}\,\varphi(t,{\bf x})\,({\cal CP})^\dag&=&{\varphi}^\ast(t,-{\bf x}) \, , \\
{\cal CPT}\,\varphi(t,{\bf x})\,({\cal CPT})^\dag&=&{\varphi}^\ast(-t,-{\bf x}) \, .
\end{eqnarray}
As we are interested to analyze possible contributions to the neutrino physics, now we are in condition to verify the behavior of the two Yukawa-type terms under {\em CP} transformation.

\subsection{Behavior of the Yukawa-type terms under {\em CP}}

Following the strategy of to looking for interactions coming from the six extra dimensions that violates the {\em CP} symmetry, we are going to study the consequences of this transformation of symmetry on the two interactions terms of Yukawa-type that appear in the Lagrangean \eqref{lagDiracfree} after the dimension reduction.

\subsubsection{Usual Yukawa-type interaction}

The usual Yukawa-type interaction terms with real $\varphi$ can be written as
\begin{equation}
 {\cal L}_{\rm Yu}=-\varphi\,\overline\psi\,\psi = -\varphi\,\psi^\dag\gamma^0\psi  \label{L-Yu}
\end{equation}
Considering the complex conjugation property to the product of Grassmann variables, or  $(\psi_1\psi_2)^\ast=-\psi_1^\ast\psi_2^\ast\,$, we have,
\begin{equation}
{\cal L}_{\rm Yu}^\ast= \varphi\,\psi^t\,{\gamma^0}^\ast\,{\psi}^\ast \, ,
\end{equation}
taking to account that $\psi^t{\gamma^0}^\ast={\overline\psi}^\ast$ and $\psi^\ast={\gamma^0}^\ast{\overline\psi}^t$, we can re-write ${\cal L}_{\rm Yu}^\ast$ as
\begin{equation}
{\cal L}_{\rm Yu}^\ast= \varphi\,{\overline\psi}^\ast\,{\gamma^0}^\ast\,{\overline\psi}^t=-\varphi\,{\overline\psi}^\ast\,\gamma^0\,{\overline\psi}^t\,. \label{L-Yu-ast}
\end{equation}
On the other hand, applying the operator ${\cal CP}$ directly to the Eq.~(\ref{L-Yu}), it follows that
\begin{equation}
{\cal CP}\,{\cal L}_{\rm Yu}\,({\cal CP})^\dag=-\varphi\,{({\overline\psi}^t)}^\dag\gamma^0\,{\overline\psi}^t=-\varphi\,{\overline\psi}^\ast\,\gamma^0\,{\overline\psi}^t\,. \label{L-Yu-CP}
\end{equation}
Comparing Eq.~(\ref{L-Yu-CP}) and Eq.~(\ref{L-Yu-ast}) it results that $({\cal CP}){\cal L}_{\rm Yu}({\cal CP})^\dag={\cal L}_{\rm Yu}^\ast$, what demonstrates that ${\cal L}_{\rm Yu}$ is invariant under {\em CP}.

\subsubsection{Pseudo-Yukawa interaction}

The interaction terms that we have called pseudo-Yukawa where $\varphi_5$  is real, have the form,
\begin{equation}
 {\cal L}_{\rm Yu5}=-i\,\varphi_5\,\overline\psi\gamma_5\psi = -i\,\varphi_5\,\psi^\dag\gamma^0\gamma_5\psi  \label{L-Yu5}
\end{equation}
By considering the complex conjugation property to the Grassmann variable product: $(\psi_1\psi_2)^\ast=-\psi_1^\ast\psi_2^\ast\,$, the above term becomes
${\cal L}_{\rm Yu5}^\ast= -i\,\varphi_5\,\psi^t\,{\gamma^0}^\ast{\gamma_5}^\ast\,\,{\psi}^\ast$,
and taking into account that $\psi^t{\gamma^0}^\ast={\overline\psi}^\ast$ ~ \mbox{and} ~ $\psi^\ast={\gamma^0}^\ast{\overline\psi}^t$, we can re-write ${\cal L}_{\rm Yu}^\ast$ as,
\begin{equation}
{\cal L}_{\rm Yu5}^\ast= -i\,\varphi_5\,{\overline\psi}^\ast\,{\gamma_5}^\ast{\gamma^0}^\ast\,{\overline\psi}^t=i\,\varphi_5\,{\overline\psi}^\ast\,\gamma^0\gamma_5\,{\overline\psi}^t\,. \label{L-Yu5-ast}
\end{equation}
So, directly applying the {\em CP} transformation on the Eq.~(\ref{L-Yu5}), it follows that,
\begin{equation}
{\cal CP}\,{\cal L}_{\rm Yu5}\,({\cal CP})^\dag=-i\,\varphi_5\,{({\overline\psi}^t)}^\dag\gamma^0\gamma_5\,{\overline\psi}^t=-i\,\varphi_5\,{\overline\psi}^\ast\,\gamma^0\gamma_5\,{\overline\psi}^t\,. \label{L-Yu5-CP}
\end{equation}
By comparing Eq.~(\ref{L-Yu5-CP}) and Eq.~(\ref{L-Yu5-ast}) it results that $({\cal CP}){\cal L}_{\rm Yu5}({\cal CP})^\dag=-{\cal L}_{\rm Yu5}^\ast$, what demonstrates that ${\cal L}_{\rm Yu5}$ violates the {\em CP} symmetry.

\acknowledgments LPGA acknowledge CNPq for the financial support. LPC, MAA and MR would like to thanks CBPF for the hospitality. MR also acknowledges the support from FAPEMIG.

\end{document}